\documentclass[fleqn,twoside]{article}
\usepackage{espcrc2}

\usepackage{graphicx}
\usepackage[figuresright]{rotating}

\newcommand{\AmS}{{\protect\the\textfont2
  A\kern-.1667em\lower.5ex\hbox{M}\kern-.125emS}}

\title{Long-Baseline Neutrino Physics in the U.S.}

\author{Sacha E. Kopp\address[UT]{Department of Physics, 
        University of Texas at Austin, \\ 
        1 University Station C1600, Austin, Texas 78712, U.S.A.}%
        \thanks{Supported by the U.S. Department of Energy, contract DE-FG03-93ER40757}}
       
\begin{document}
\begin{abstract}
  Long-baseline neutrino oscillation physics in the U.S. is centered at the Fermi National Accelerator Laboratory (FNAL), in particular at the Neutrinos at the Main Injector (NuMI) beamline commissioned in 2004-2005.  Already, the MINOS experiment has published its first results confirming the disappearance of $\nu_\mu$'s across a 735~km baseline.  The forthcoming NO$\nu$A experiment will search for the transition $\nu_\mu\rightarrow\nu_e$ and use this transition to understand the mass heirarchy of neutrinos.  These, as well as other conceptual ideas for future experiments using the NuMI beam, will be discussed.  The turn-on of the NuMI facility has been positive, with over 310~kW beam power achieved.  Plans for increasing the beam intensity once the Main Injector accelerator is fully-dedicated to the neutrino program will be presented.\vspace{1pc}
\end{abstract}

\maketitle

\section{Introduction}

A staged program of neutrino oscillation experiments is forseen using a new beam line facility, the ``Neutrinos at the Main Injector'' (NuMI) \cite{numitdr}, at the Fermi National Accelerator Laboratory.  The first experiment, MINOS \cite{minos}, will perform definitive spectrum measurements which demonstrate the effect of $\nu_\mu$ oscillations.  The experiment has run since early 2005 and published a first result \cite{minos-prl} based upon $1.3\times10^{20}$~protons-on-target, approximately 1/10$^{th}$ the eventual exposure of the experiment.  The second, NO$\nu$A \cite{nova}, has been approved to explore the phenomenon of CP violation in the transition $\nu_\mu\rightarrow\nu_e$ and possibly resolve the mass heirarchy of neutrinos.  A third, MINER$\nu$A \cite{minerva}, is an experiment 1 km from the NuMI target to perform neutrino cross section measurements essential for long-baseline neutrino oscillation searches.

\section{NuMI Intensity and Upgrades}

\begin{figure*}[htb]
\vspace{9pt}
  \includegraphics[width=5.8in]{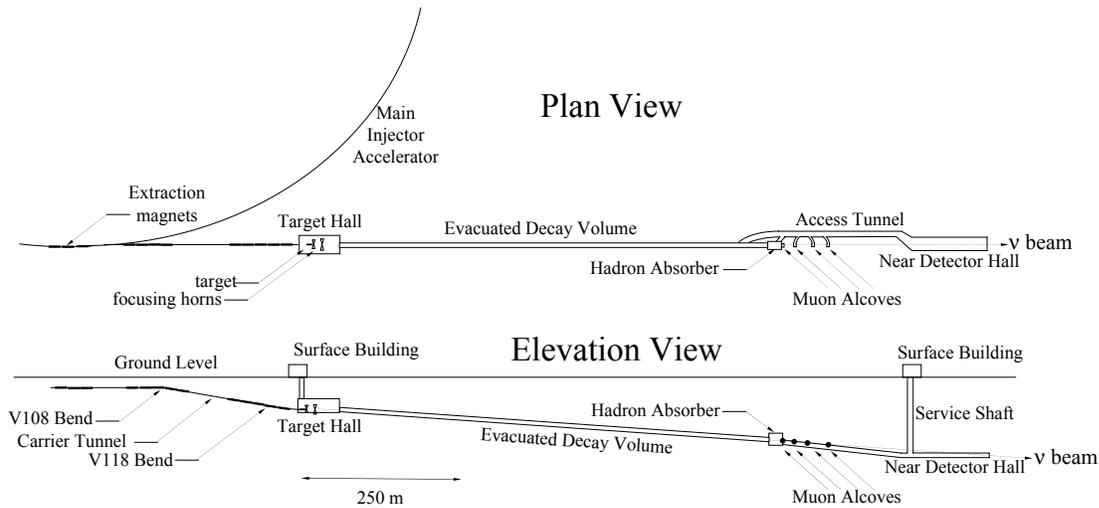}
\caption{Plan and elevation view of the NuMI beam at FNAL.  }
\label{fig:numi}
\end{figure*}

NuMI is a tertiary beam resulting from the decays of pion and kaon secondaries produced in the NuMI target. Protons of 120 GeV are fast-extracted (spill duration $\sim10\mu$sec) from the Main Injector (MI) accelerator and bent downward by 58~mrad toward Soudan, MN (see Figure~\ref{fig:numi}).  The beam line is designed to accept $4\times10^{13}$~protons-per-pulse~(ppp).  The beam line can be flexibly configured to achieve a variety of neutrino energies \cite{flexybeam}.  
To date, the average (peak) beam power has been 230~kW (310~kW), and a three-phase plan has been approved to upgrade the accelerator complex and NuMI line to increase this intensity to 430~kW by 2009, then 700~kW by 2012, then 1200~kW by 2013.  

The MI is fed multiple batches from the 8~GeV Booster accelerator.  The Booster can deliver $(4-5)\times10^{12}~p$/batch.  During normal operations, the MI beam is shared:  2 of 7 accelerated Booster batches are extracted to make antiprotons for the Tevatron collider, while the remaining 5 are extracted to the NuMI target.  

The first phase of the intensity upgrades takes advantage of the fast (67~ms) cycle time to the Booster to inject as many as 11 batches of protons into the MI.
 Pairs of batches will be coalesced by a process of ``slip-stacking'' \cite{kiyomi} into 6 double-batches around the MI circumference, and this beam accelerated to 120~GeV.  Already, MINOS has received a 20\% increase in beam power by slip-stacking of 2 batches and accelerating these along with 5 other single batches, and tests of 11-batch stacking have achieved as much as 3$\times10^{13}$~ppp, which will improve as the coalescing and capture process is better understood.  This phase should be fully-implemented in 2007 and achieve 430~kW.  

The second phase will use the 8~GeV Recycler ring in the MI tunnel as a pre-injector.  The Recycler, which has the same circumference as the MI, will be used to pre-load and slip-stack 11 Booster batches, then perform a fast transfer to the MI for acceleration.  This reduces the MI cycle time from 2.2~sec. to 1.3~sec., increasing the beam power to NuMI to 700~kW (see Table~\ref{table:power}).  This phase is to be implemented at the end-of-Tevatron shutdown in 2009, and has been approved by the Laboratory.

\begin{figure}[htb]
\vspace{9pt}
  \includegraphics[width=3in]{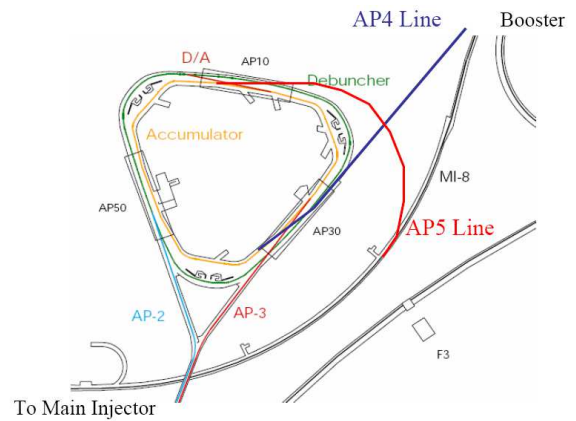}
\vskip -1.cm
\caption{Plan view of the FNAL Accumulator ring, with new AP4 injection line from the Booster and AP5 extraction line to the MI.}
\label{fig:accum}
\vskip -1.cm
\end{figure}

The third phase, which is still in its conceptual stages, calls for use of the Accumulator, currently utilized for antiproton beam accumulation, as a proton beam accumulator which will permit 3 Booster batches to be coalesced in one Booster-equivalent circumference.  Six such Accumulator batches can be injected into the Recycler, then the Main Injector.  The plan calls for new injection and extraction lines to be built for the Accumulator (see Figure~\ref{fig:accum}).  As shown in Table~\ref{table:power}, the anticipated beam power to NuMI is then 1.2~MW.

Phases II and III require upgrades to the NuMI line.  These include the substitution of a graphite target with alternative water cooling schemes capable of withstanding the higher beam power and also additional cooling in the target station, decay tunnel, and beam stop.  The new target design is facilitated by the fact that in the future the NuMI beam will operate at higher on-axis neutrino energy (which will obtain an ideal 1.8~GeV off-axis for the NO$\nu$A experiment, see below) and the target will no longer be cantilevered inside the first focusing horn.

\begin{table*}[htb]
\caption{Present and projected NuMI beam power following 3 upgrades.}
\label{table:power}
\newcommand{\m}{\hphantom{$-$}}
\newcommand{\cc}[1]{\multicolumn{1}{c}{#1}}
\renewcommand{\tabcolsep}{2pc} 
\renewcommand{\arraystretch}{1.2} 
\begin{tabular}{@{}llll}
\hline
                                      & Protons/pulse & Cycle time & Power  \\
\hline
{Current Complex}   &               &            &        \\
$\cdot$ Shared Beam     & 25$\times10^{12}$ & 2.4 s & 200 kW \\
$\cdot$ NuMI Alone     & 33$\times10^{12}$ & 2.0 s & 310 kW \\
{Phase I (2009)$^{(a)}$}   &               &            &        \\
$\cdot$ Shared Beam     & 37$\times10^{12}$ & 2.2 s & 320 kW \\
$\cdot$ NuMI Alone     & 49$\times10^{12}$ & 2.2 s & 430 kW \\
{Phase II$-$Recycler (2012)$^{(a)}$}   & 49$\times10^{12}$ & 1.3 s & 700 kW \\
{Phase III$-$Accumulator (2013)$^{(b)}$}   & 83$\times10^{12}$ & 1.3 s & 1200 kW \\
\hline
 & $^{(a)}$Approved &\multicolumn{2}{l}{$^{(b)}$Still at conceptual stage.}
\end{tabular}\\[2pt]
\end{table*}

\section{MINOS}

The Main Injector Neutrino Oscillation Search (MINOS) \cite{minos} began data-taking in 2005 and has accumulated $1.8\times10^{20}$~POT.  The first results \cite{minos-prl} based on 1.2$\times10^{20}$~POT were covered by B. Rebel at this workshop.  MINOS is a two-detector experiment, with a 980~ton detector located 1~km from the NuMI target on site at FNAL (see Figure~\ref{fig:numi}) to measure the neutrio flux directly, and a second, 5400~ton detector located in the Soudan mine in Minnesota, at a distance 735~km.  The spectrum of neutrinos arriving in the far detector is studied to investigate whether neutrinos oscillate, decay, or otherwise disappear in flight.   

\begin{figure}[htb]
  \includegraphics[width=3.1in]{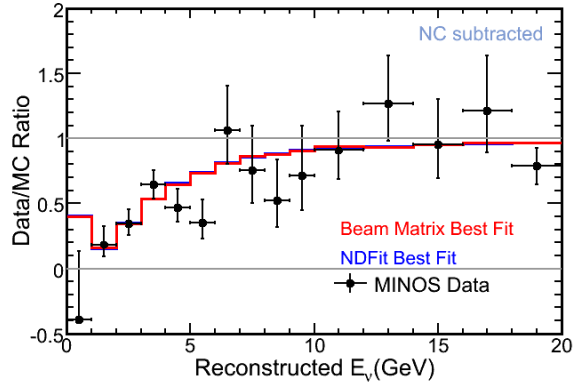}
\vskip -1.cm
\caption{First result from the MINOS experiment \cite{minos-prl}.  A $>6\sigma$ deficit of neutrinos is observed in the far detector in Minnesota.  }
\label{fig:minos-result}
\end{figure}

MINOS has investigated the rate of charged-current $\nu_\mu$ interactions in the near and far detectors.  Oscillations would result in an energy-dependent depletion of these interactions.  The results of the first 1.2$\times10^{20}$~POT are shown in Figure~\ref{fig:minos-result}.  Following commissioning of the beam line, the experiment acquired several data samples at different neutrino energies to study the beam spectrum and neutrino interactions in the near detector.  Having thus tested the calculated spectrum of the neutrino flux, the Monte Carlo could be better trusted to describe any differences in the neutrino energy spectra in the near and far detectors.  
The energy spectrum observed in the near detector was multiplied by an extrapolation factor \cite{para} which corrects the near spectrum to the expected spectrum (in the absence of oscillations) at the far detector.  A deficit of $\nu_\mu$'s with $E<6$~GeV of $>6\sigma$ is observed.  If interpreted as neutrino oscillations, the data indicate

\begin{equation}
\Delta m^2_{32}=(2.74^{+0.44}_{-0.26})\times10^{-3}~\mbox{eV}^2 
\end{equation}
with the mixing angle consistent with maximal mixing:  $\sin^2(2\theta_{23})>0.87$.  The expected resolution is $\delta(\Delta m_{32}^2)<10^{-4}~\mbox{eV}^2$ at $16\times10^{20}$~POT.

\begin{figure}[tb]
  \includegraphics[width=2.9in]{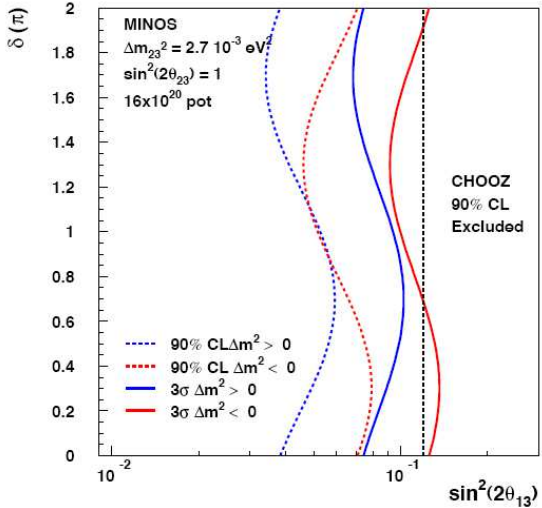}
\vskip -1.cm
\caption{Expected sensitivity of the MINOS experiment to the appearance of $\nu_e$'s in its $\nu_\mu$ beam.    }
\label{fig:minos-nue}
\end{figure}

With further statistics, the experiment hopes to observe or place stringent limits on the transition $\nu_\mu\rightarrow\nu_e$.  As shown in Figure~\ref{fig:minos-nue}, the potential for 3$\sigma$ discovery depends upon the values of the CP-violating phase $\delta$ and the sign of $\delta m_{32}^2$.  

\section{MINER$\nu$A}

\begin{figure}[tb]
\vspace{9pt}
  \includegraphics[width=3.in]{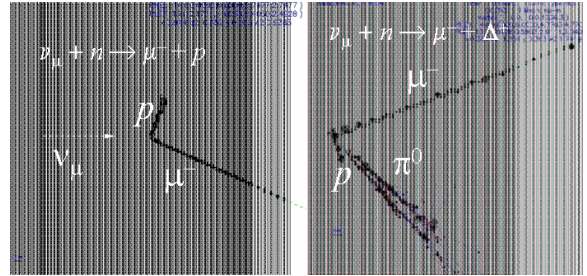}
\vskip -0.5cm
\caption{Display of two simulated neutrino interactions in the MINER$\nu$A detector:  (left) quasielastic scattering, (right) resonant production, with the $\Delta^+\rightarrow p\pi^0$ visible in the detector. }
\label{fig:minerva-events}
\end{figure}

The MINER$\nu$A experiment is a fine-grained scintillator detector like the KEK SciBar detector, but located in the NuMI detector hall 1~km from the NuMI target.  It features 196 alternating planes of $U$ and $V$ strips, with each strip consisting of $1.7\times3.3~\mbox{cm}^2$ triangular solid extruded scintillator read out into a multi-anode PMT via scintillating fiber.  The finer pitch, overlapping triangular strips gives excellent position resolution to reconstruct final states in neutrino-nucleus interactions.  A coarse outer calorimeter is used for shower containment, and muon momentum analysis is performed in the downstream MINOS near detector.  The ability of the detector to perform detailed analyses of neutrino final states may be seen in Figure~\ref{fig:minerva-events}.

Uncertainties in neutrino cross sections are substantial at low energy.  Of particular note are the large uncertainties in the MINOS oscillation result due to such cross section uncertainties \cite{minos-prl}:  pion reabsorption by the struck nucleus can alter the apparent final state multiplicity and visible energy.  Such nuclear effects will be studied using the 18 nuclear target sheets in MINER$\nu$A.

\section{NO$\nu$A}

The NuMI Off-Axis $\nu_e$ Appearance (NO$\nu$A) experiment has been approved by the Laboratory to search for and use the transition $\nu_\mu\rightarrow\nu_e$ to study CP violation in the neutrino sector.  

\begin{figure}[t]
\vspace{9pt}
\vskip -0.5cm
  \includegraphics[width=2.5in]{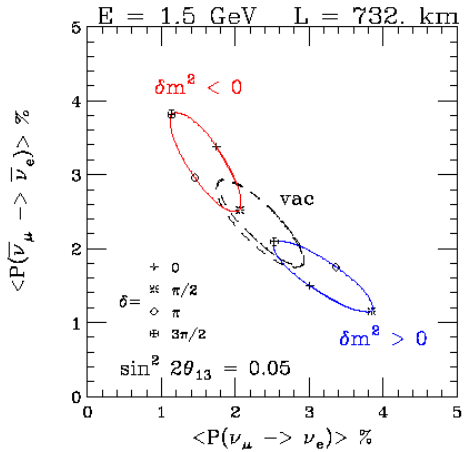}
\vskip -1.cm
\caption{Probability for the transition $\nu_\mu\rightarrow\nu_e$ and $\overline{\nu}_\mu\rightarrow\overline{\nu}_e$ for 735~km baseline and $E_\nu$=1.5~GeV as a function of the CP violating phase $\delta$ under two assumptions for the mass ordering.  Taken from \cite{nova}.}
\label{fig:nue-matter}
\end{figure}

\begin{figure}[h]
\vspace{9pt}
\vskip -0.5cm
  \includegraphics[width=2.6in]{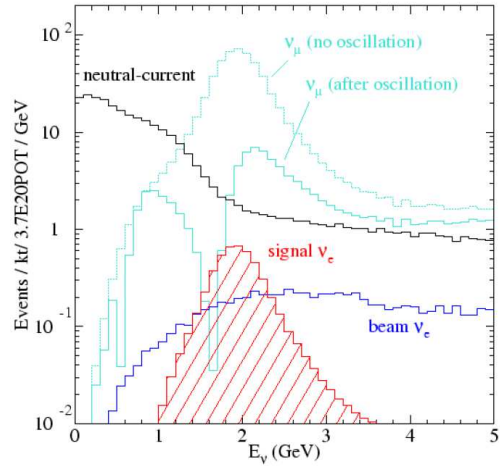}
\vskip -1.cm
\caption{The charged-current $\nu_\mu$ spectrum expected in NO$\nu$A, along with the expected $\nu_e$ signal from oscillations with $\sin^2 2\theta_{13}=0.1$ and $\nu_e$ contamination from the beam.  Also shown is the visible energy from neutral current $\nu_\mu$ interactions before any rejection criteria are applied.  }
\label{fig:nova-spectrum}
\end{figure}

The NuMI long baseline and higher neutrino energy results in a splitting between the transition probability for $\nu_\mu\rightarrow\nu_e$ and $\overline{\nu}_\mu\rightarrow\overline{\nu}_e$, as shown in Figure~\ref{fig:nue-matter}.  This arises due to the matter effect, and results in an asymmetry even in the absence of CP violation.  This fact makes the NO$\nu$A experiment quite complementary to the JPARC T2K experiment, which will operate at lower $E_\nu$ and shorter baseline:  JPARC will measure the transition probability for $\nu_\mu\rightarrow\nu_e$, which measures the mixing angle $\theta_{13}$ in the neutrino mixing matrix, and the NO$\nu$A experiment could study both the $\nu$ and $\overline{\nu}$ process to understand the mass heirarchy of the neutrino states, and possibly also measure the CP phase $\delta$.

\begin{figure}[tb]
\vspace{9pt}
  \includegraphics[width=2.9in]{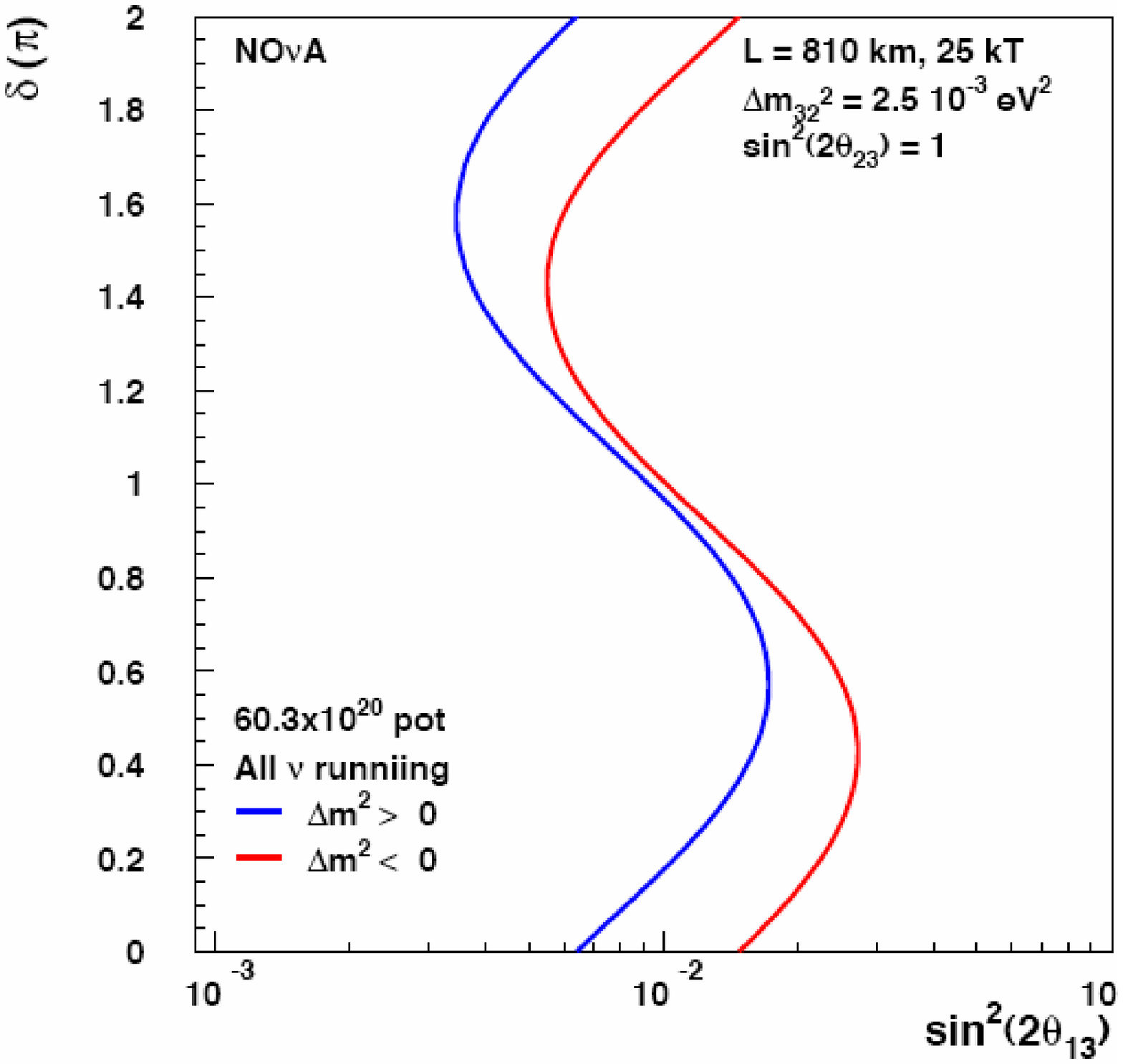}
  \includegraphics[width=2.9in]{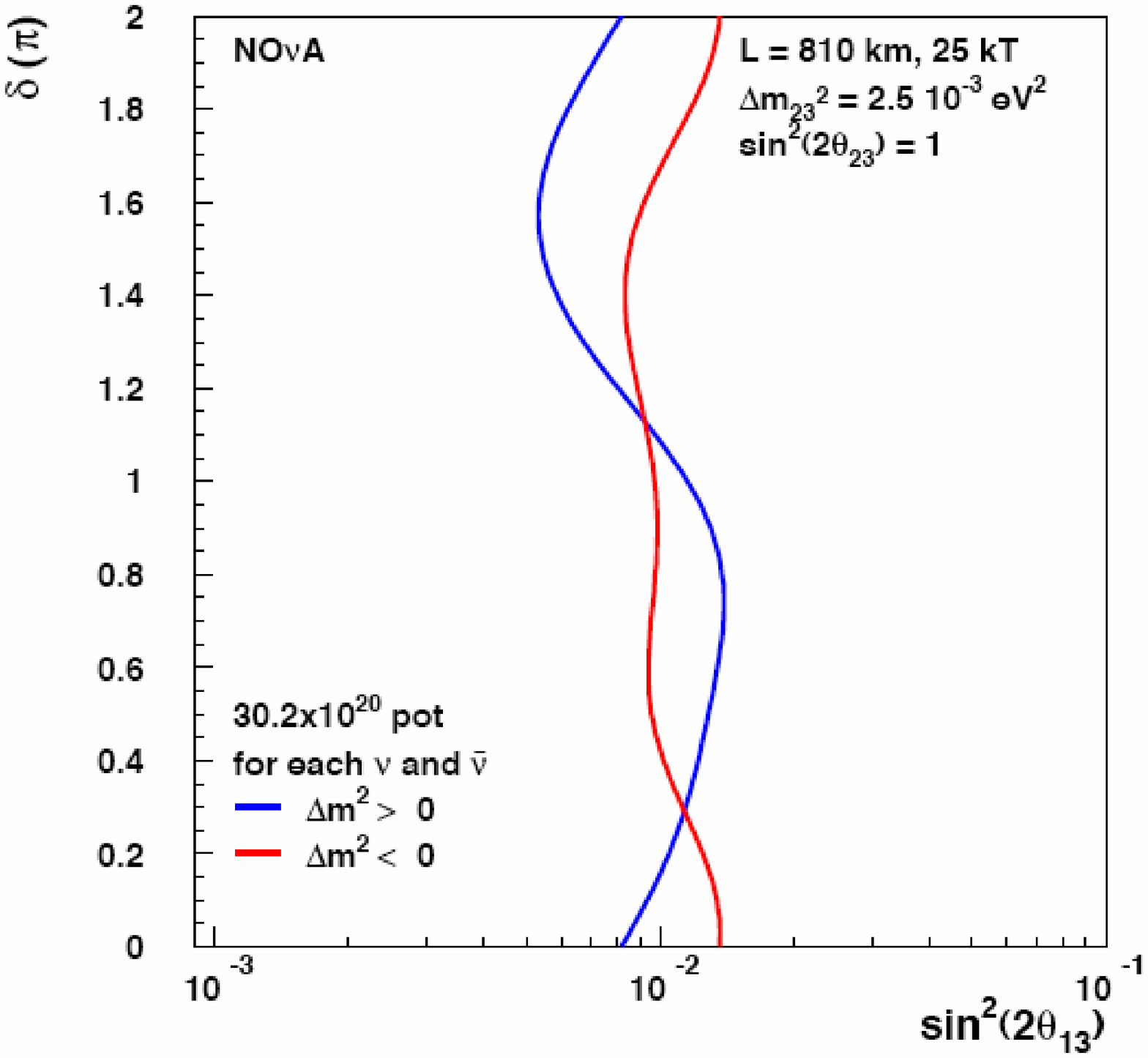}
\vskip -1.cm
\caption{Curves showing the $3\sigma$ discovery potential of the NO$\nu$A experiment to observe the transition $\nu_\mu\rightarrow\nu_e$ after an exposure of $60\times10^{20}$~POT.  The upper plot is for an all-neutrino exposure, the lower for an equal admixture of $\nu$ and $\overline{\nu}$ running.}
\label{fig:nova-sensitivity}
\end{figure}

NO$\nu$A must look for the appearance of charged-current $\nu_e$ interactions in a $\nu_\mu$ beam, a challenge because because of the small transition probability and the prevalence of showers from neutral current $\nu_\mu$ interactions which can mimic the signal.  As shown in Figure~\ref{fig:nova-spectrum}, the $\nu_\mu$ off-axis spectrum can be centered at the oscillation maximum for $\nu_\mu\rightarrow\nu_e$, and the putative signal for $\sin^2 2\theta_{13}=0.1$ is significant over the intrinsic $\nu_e$ in the beam which arises primarily from muon decays in flight.  The challenge for the experiment will be the reduction of neutral current $\nu_\mu$ interactions, which can mimic the showers of charged-current $\nu_e$ events.  

To reduce neutral current backgrounds, the NO$\nu$A detector is a fine-grained totally-active segmented calorimeter, consisting of alternating horizontally- and vertically-oriented planes of scintillator strips, 15.7~m in length and 6~cm in depth by 3.9~cm transverse width.  The scintillator strips are made using extruded PVC shells filled with liquid scintillator.  The two ends of a looped scintillating fiber which runs the 15.7~m length of the strip brings the light out to an APD.  The light output for a minimum ionizing particle is 20~photo-electrons, to be compared with 2~photo-electrons noise from the APD.  The detector will have 1654 planes of 384 strips, for a total mass of 25~kton.  

Figure~\ref{fig:nova-sensitivity} shows the capability of the NO$\nu$A experiment to observe the $\nu_\mu\rightarrow\nu_e$ transition with 3$\sigma$ significance after an exposure of $60\times10^{20}$~POT (equivalent to 7~years of Phase-III running).  The curves are calculated using values derived from a full simulation of neutrino interactions in the NO$\nu$A detector, namely 23\% efficiency for identifying a $\nu_e$ charged current interaction, as well as a rejection factor of 7$\times10^{-4}$ (1.3$\times10^{-3}$) for charged current (neutral current) $\nu_\mu$'s.  
Running the NuMI beam in both $\nu$ and $\overline{\nu}$ mode makes the discovery potential of the experiment less dependent upon the CP phase and mass heirarchy.

\section{Future Initiatives}

Two longer-term design studies are underway in the U.S.  The first would consider the feasibility of building a high-resolution liquid Argon detector of order 50~kton to study $\nu_\mu\rightarrow\nu_e$ transitions in the NuMI beam \cite{flare}.  The design would build upon previous LAr experience, housing the detector in a massive cryogenic system as is used for liquid natural gas.

The second is a BNL-FNAL study \cite{diwan} to send a neutrino beam to DUSEL.  If the beam were initiated at FNAL, the Main Injector complex could serve as its source, but a new neutrino line would have to be built which points in the direction of DUSEL.  The proposal is for an on-axis wide-band neutrino beam.  At the FNAL site, such a beam line could be at most 400~m in length.

\section{Summary}

Neutrino physics in the U.S. has enjoyed a renaissance with the commissioning of the NuMI line at FNAL.  The MINOS experiment has already published a competitive measurement of $\Delta m^2_{32}$, and still looks forward to collecting 10 times more data.  In the future, we may look forward to oscillation and scattering results.

The NO$\nu$A experiment is an ambitious effort to use $\nu_\mu\rightarrow\nu_e$ to study the $\nu$ mass heirarchy and search for CP violation in the lepton sector.  It complements the shorter-baseline experiment at J-PARC or a reactor disappearance experiment.  

The NuMI line is commissioned and performing well.  As the Tevatron collider ramps down, Fermilab will be a dedicated neutrino facility, and the accelerator complex will be re-commissioned toward higher beam power.  

I wish to acknowledge my colleagues on the NuMI facility and MINOS, NO$\nu$A, and MINER$\nu$A experiments.  R. Zwaska, A. Marchionni, M. Martens, and N. Grossman kindly provided information on the proton source upgrades.


\begin{thebibliography}{9}
\bibitem{numitdr} J. Hylen et al., Fermilab-TM-2018, 1997.
\bibitem{minos} D.G. Michael {\it et al}, ``The MINOS Detector,'' submitted to NIM.
\bibitem{minos-prl} D.G. Michael {\it et al.}, Phys. Rev. Lett. {\bf 97}:191801 (2006).
\bibitem{nova} NOvA Collaboration, {\tt hep-ex/0210005} (2002), G. Feldman and M. Messier, spokespersons.
\bibitem{minerva} MINER$\nu$A Collaboration, {\tt hep-ex/0405002}  (2004), J. Morfin and K. McFarland, spokespersons.
\bibitem{flexybeam} M. Kostin {\it et al.}, Fermilab-TM-2353(AD) (2001).
\bibitem{kiyomi} K. Koba {\it et al.}, Proc. Part. Accel. Conf. (2003), p. 1736.
\bibitem{para} A. Para and M. Szleper, {\tt arXiv:hep-ex/0110001}.
\bibitem{flare} L. Bartoszek {\it et al.}, {\tt arXiv:hep-ex/0408121}.
\bibitem{diwan} M. Diwan {\it et al.}, {\tt arXiv:hep-ex/0608023}.
\end{thebibliography}
\end{document}